\documentclass[a4paper,12pt]{article}
\usepackage[a4paper,text={16.8cm,22.4cm}]{geometry}
\usepackage{amsmath,amssymb,bm,graphicx,color,psfrag}
\allowdisplaybreaks 
\def\rd{\mathrm{d}}

\def\eps{\epsilon}

\begin{document}

\begin{titlepage}

\begin{flushright}
January 26, 2012
\end{flushright}

\vspace{0.7cm}
\begin{center}
\Large\bf
\boldmath
NNLO soft function for electroweak boson production at large transverse momentum
\unboldmath
\end{center}

\vspace{0.2cm}
\begin{center}
{\sc Thomas Becher, Guido Bell and Stefanie Marti}\\
\vspace{0.4cm}
{\sl 
Albert Einstein Center for Fundamental Physics\\
Institute for Theoretical Physics \\
University of Bern\\
Sidlerstrasse 5, 3012 Bern, Switzerland}
\end{center}

\vspace{1.0cm}
\begin{abstract}
\vspace{0.2cm}
\noindent 
The  soft function relevant  for the production of an electroweak boson ($\gamma$, $W$, $Z$ or $H$) with large transverse momentum at a hadron collider is computed at next-to-next-to-leading order. This is the first two-loop computation of a soft function involving three light-cone directions. With the result, the threshold resummation for these processes can now be performed at next-to-next-to-next-to-leading logarithmic accuracy.
\end{abstract}
\vfil

\end{titlepage}

\section{Introduction\label{introduction}}

The production of an electroweak boson, followed by its decay to leptons, is one of the most basic hard-scattering processes at hadron colliders. With more than $5\,{\rm fb}^{-1}$ of data, the LHC experiments have now recorded millions of $Z$  and $W$ bosons decaying into lepton pairs, which allows for precision measurements, even at large transverse momentum $p_T$ of the boson. Given that $Z$'s and $W$'s at high $p_T$ provide an important background to new physics searches and are used to calibrate jet energy scales, it is important to have good theoretical control of the cross section in this region. The complete ${\mathcal O}(\alpha_s^2)$ corrections to vector boson production are known and have been implemented into numerical codes which allow for arbitrary cuts on the final state \cite{Melnikov:2006di,Melnikov:2006kv,Catani:2009sm,Gavin:2010az}. However, since the $p_T$-spectrum starts at ${\mathcal O}(\alpha_s)$, these codes only provide the next-to-leading-order (NLO) corrections for boson production at large transverse momentum. There is an  ongoing effort to  evaluate electroweak boson production in association with a jet to NNLO. The necessary two-loop results for the production of photons, $W$'s and $Z$'s are known for some time \cite{Garland:2001tf,Garland:2002ak,Gehrmann:2011ab}. Very recently, also the two-loop results for Higgs production in association with a jet were presented \cite{Gehrmann:2011aa}. To obtain the transverse momentum spectrum at NNLO, these virtual corrections have to be combined with the real-emission corrections, which is a difficult task because of the very singular nature of the individual contributions.

While the real-emission corrections are complicated in general, they simplify close to the partonic threshold. In this region, the hadronic final state consists of a single, low-mass jet and all radiation must  either be soft, or collinear to the jet or the incoming hadrons. In this situation, the partonic cross section factorizes into a hard function times the convolution of a jet function with a soft function:
\begin{equation}\label{factform}
\hat{\sigma}  =  H \cdot  J \otimes S \, .
\end{equation}
This factorization holds channel by channel, but with different hard, jet and soft functions.
For $\gamma$, $Z$ or $W$ production, there are two channels, the annihilation channel $ q + \bar{q} \to g + V$ and the Compton process $ q + g \to q + V$. For Higgs production, the two relevant channels are $g + g \to g + H$ and $q + g \to q + H$. A detailed derivation of the factorization formula in Soft-Collinear Effective Theory (SCET)  \cite{Bauer:2000yr,Bauer:2001yt,Beneke:2002ph} and one-loop results for various ingredients were given in \cite{Becher:2009th}. By now, almost all of the ingredients to obtain the threshold cross section to NNLO are known. The hard function $H$ contains the virtual corrections, and can be obtained from the results in \cite{Garland:2001tf,Garland:2002ak,Gehrmann:2011aa}. The two-loop jet functions $J$ for quark and gluon jets were evaluated in \cite{Becher:2006qw,Becher:2010pd}. The only missing ingredient is the NNLO soft function $S$ which is computed in the present paper. 

The scale associated with the soft radiation is lower than the scale $p_T$ which is relevant for the hard function. As a consequence, the cross section contains perturbative logarithms of  scale ratios which should be resummed. At next-to-leading logarithmic (NLL) accuracy, this was achieved in \cite{Kidonakis:1999ur,Kidonakis:2003xm,Gonsalves:2005ng}. In the papers \cite{Becher:2009th,Becher:2011fc, KidonakisNew} the resummation was recently extended to NNLL. Using the results of \cite{Becher:2009cu,Gardi:2009qi,Becher:2009qa}, all the anomalous dimensions necessary for N$^3$LL resummation were derived in \cite{Becher:2009th}. With the soft function computed in the present paper, the resummation can thus now be performed at N$^3$LL accuracy.
 
\begin{figure}
\begin{center}
\begin{tabular}{ccccc}
\includegraphics[height=3.5cm]{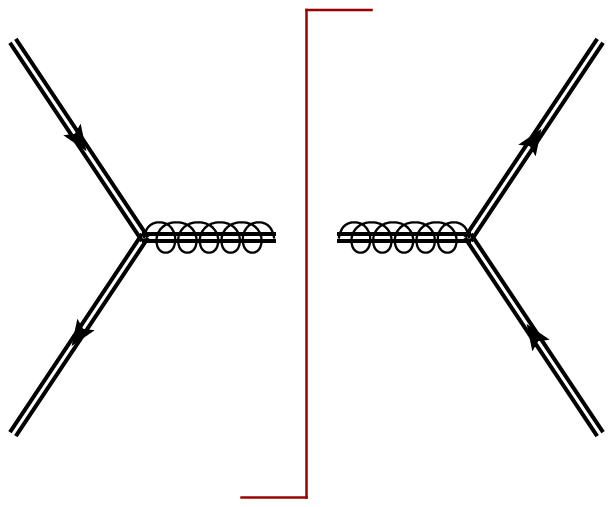} && \includegraphics[height=3.5cm]{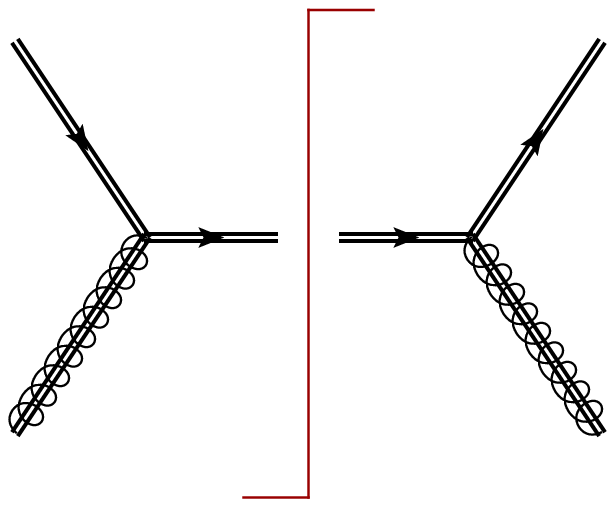} && \includegraphics[height=3.5cm]{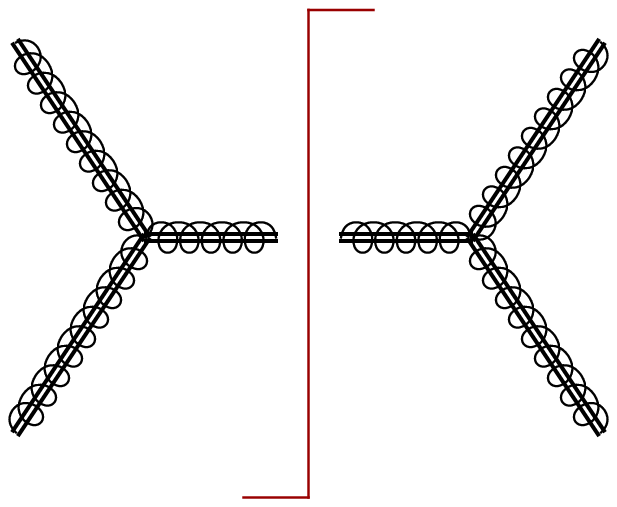} \\
$q\,\bar q \to g \, V $ & & $q\, g \to q \, V $ &&  \\
 & & $q\, g \to q \, H $ && $g\, g \to g \, H $ \\
\end{tabular}
\end{center}
\vspace{-4mm}
\caption{\label{fig:star}
The three Wilson-line configurations relevant for electroweak boson production. \label{fig:channels}}
\end{figure}

The soft emissions are described by Wilson lines along the directions $n_i^\mu = p_i^\mu/E_i$  of the partons which participate in the hard-scattering process. To be able to derive results independent of the color representation of the partons, we use the color-space formalism introduced in \cite{Catani:1996jh,Catani:1996vz}. The Wilson line for a particle in a color representation with generators $\bm{T}_i^a$ is defined as the path-ordered exponential
\begin{equation}\label{wilsondef}
\bm{S}_i(x) = {\bf P} \exp\left( i g_s \int_{-\infty}^0 dt\,  n_i\cdot A^a(x+t \,n_i)\, \bm{T}_i^a   \right) \, .
\end{equation}
If the particle is a gluon carrying a color index $c$, we have $(\bm{T}^a)_{bc}=-i\, f_{abc}$, and for an outgoing quark (or incoming anti-quark) with index $\beta$ the generator is $(\bm{T}^a)_{\alpha \beta} = t^a_{\alpha\beta}$. For an incoming quark (or outgoing anti-quark) with index $\beta$, the generator is $(\bm{T}^a)_{\alpha \beta} = -t^a_{\beta\alpha}$, which translates into anti-path ordering in (\ref{wilsondef}) so that ${\bf S}_{\bar{q}}(x)={\bf S}^\dagger_q(x)$. 

The specific Wilson line configurations relevant for electroweak boson production in the different partonic channels are shown in Figure \ref{fig:channels}. We will compute the soft function for the general case, where the initial-state Wilson lines are in representations $1$ and $2$, while the final-state Wilson line is in the representation $J$. We denote the directions of the associated Wilson lines by $n_1$, $n_2$ and $n_J$, respectively. The soft function then has the form
\begin{equation}\label{softdef}
 \bm{S}(\omega) = \sum_X  \langle 0|  \bm{S}^\dagger_1 \, \bm{S}^\dagger_2\, \bm{S}^\dagger_J | X \rangle  \langle X | \bm{S}_1 \, \bm{S}_2 \, \bm{S}_J |0\rangle \, \delta(\omega-n_J\cdot p_X)\,,
\end{equation}
with $\bm{S}_i\equiv \bm{S}_i(0)$. It measures the probability for soft emissions from the three Wilson lines with given momentum component $\omega=n_J\cdot p_X$ along the jet direction. This component is relevant since the invariant mass of the final-state jet is given by the momentum $p_c\approx E_J\, n_J^\mu$ of the particles collinear to the jet and the momentum $p_X$ of the soft particles as
\begin{equation}
M_J^2 = (p_c+ p_X)^2 \approx p_c^2+ 2 E_J\, n_J\cdot p_X\,.
\end{equation}

The color indices of the Wilson lines $\bm{S}_i$ are contracted with the color indices of the hard-scattering amplitude, while the conjugate Wilson lines $\bm{S}^\dagger_i$ are contracted with the indices of the complex conjugate amplitude.  These amplitudes collect the virtual corrections to the scattering process and their product is the hard function $H$  in (\ref{factform}). However, the hard-scattering process involves only three partons and the color algebra of these amplitudes is diagonal \cite{Catani:1996jh}. As a consequence, the soft function reduces to a number times the trivial color structure $ \bm{S}(\omega)=  S(\omega) \,\bm{1}$ when applied to the hard function in a given channel. The function $S(\omega)$ depends on the channel through the Casimir invariants associated with the representations of the three Wilson lines.

By now, a number of NNLO results for soft functions involving two Wilson lines can be found in the literature, many of them appeared during the past year \cite{Belitsky:1998tc,Becher:2005pd,Kelley:2011ng,Monni:2011gb,Hornig:2011iu,Li:2011zp,Kelley:2011aa}. For the case of multiple Wilson lines, only the anomalous dimensions were analyzed, first in the massless case \cite{Aybat:2006wq,Aybat:2006mz} and more recently also in the massive case \cite{Mitov:2009sv,Ferroglia:2009ep,Ferroglia:2009ii}. In the massless case, the two-loop anomalous dimension was found to be proportional to the one-loop result. What came as a surprise at the time, is now understood to be a consequence of the strong constraints to which soft anomalous dimensions are subject \cite{Becher:2009cu,Gardi:2009qi,Becher:2009qa, Dixon:2009ur, DelDuca:2011xm,DelDuca:2011ae}. In our paper, we present the first full two-loop result for a soft function with three Wilson lines. Interestingly, these constraints, in particular the invariance of Wilson lines under a rescaling of the reference vector, drastically simplify our computation. As we will show below, all diagrams with attachments to the jet Wilson line  vanish and the only non-zero diagrams are the ones which are present already in the case with two Wilson lines. 

In the next section, we turn to the evaluation of  the diagrams. We first demonstrate that the soft function is color diagonal and that only a very limited set of diagrams contribute. We then discuss how the corresponding integrals can be evaluated and give the bare result for the soft function. In Section \ref{renorm}, we  perform the renormalization and present our final result.  Technical details and explicit results for all loop integrals needed in the computation are given in three appendices.

\section{NNLO calculation\label{nnlo}}

Before proceeding to the computation of the diagrams, we now demonstrate that the color structure of the soft function is diagonal and that the dependence on the light-cone vectors is completely determined by rescaling invariance. The color indices of the soft Wilson lines are contracted with the hard-scattering amplitudes. In color-space notation, the hard amplitude for the annihilation channel can be written in the form
\begin{equation}
 | C_{q \bar{q} \to g V}\, \rangle_{\rm color} =  t_{\beta\alpha}^a\, C_{q\bar{q}\to g V}\,
\end{equation}
where $\alpha$, $\beta$ and $a$ are the color indices of the quark, anti-quark and gluon, respectively, and $C_{q\bar{q}\to g V}$ is a function of the momenta and spins only. We have added a subscript to the vector $ | C_{q \bar{q} \to g V}\, \rangle$ to distinguish the color states of the hard function from the state vectors of the Hilbert space of the soft partons.  The relevant color structure for Higgs production in the gluon channel is
\begin{equation}
 | C_{g g \to g H}\, \rangle_{\rm color} = i f_{abc} \,C_{gg\to g H}\,.
\end{equation}
On general grounds, also the color structure $d_{abc}$ could arise, but at least up to two-loop order  this does not happen \cite{Gehrmann:2011aa}. 

 \begin{figure}[t!h]
 \begin{center}
 \begin{tabular}{ccccc}
 \psfrag{n1}[]{\small $n_1$}
 \psfrag{n2}[]{\small $n_2$}
 \psfrag{nJ}[lB]{\small $n_J$}
 \includegraphics[height=0.19\textwidth]{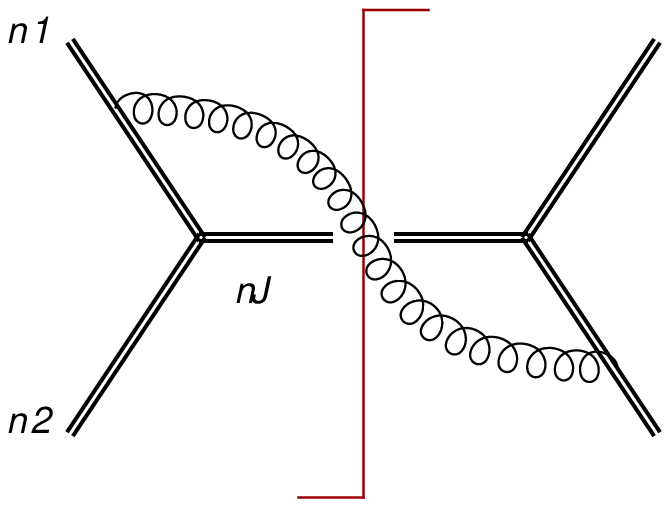} &&\includegraphics[height=0.19\textwidth]{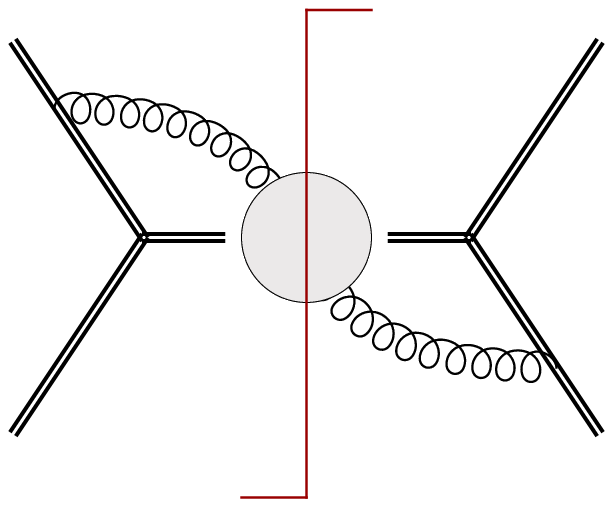}  &&\includegraphics[height=0.19\textwidth]{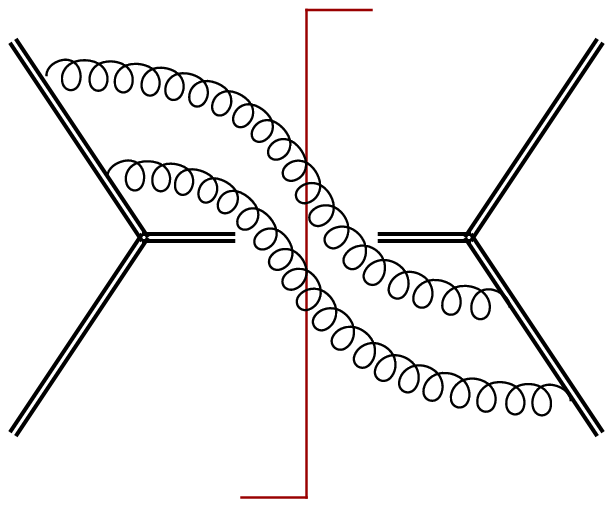} \\[0.01cm]
 $D_1$&&$D_2$&&$D_3$\\[0.6cm]
  \includegraphics[height=0.19\textwidth]{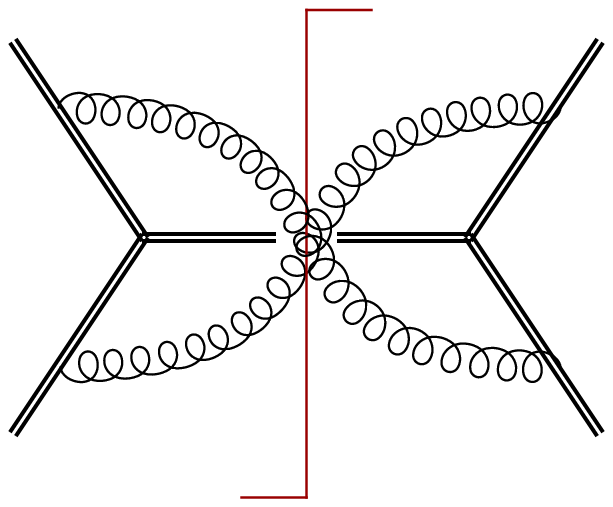} &&\includegraphics[height=0.19\textwidth]{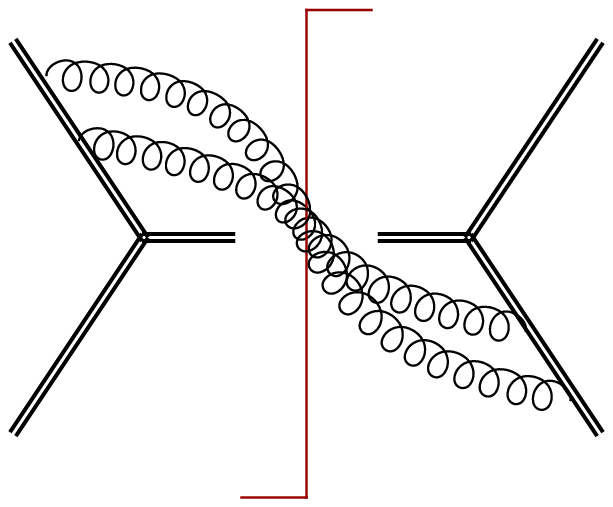}  &&\includegraphics[height=0.19\textwidth]{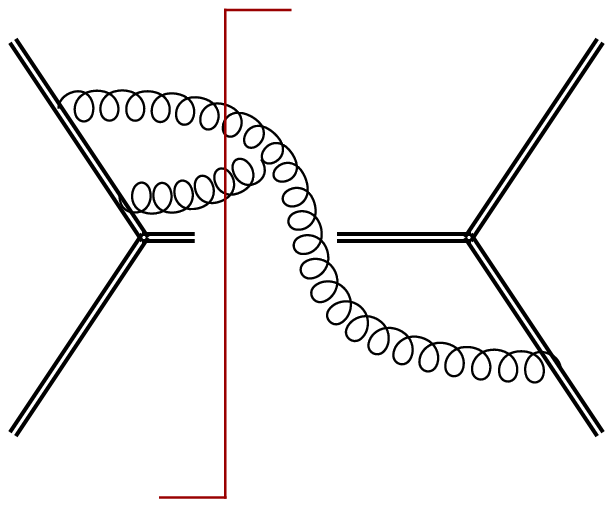} \\[0.01cm]
 $D_4$&&$D_5$&&$D_6$\\[0.6cm]
 \includegraphics[height=0.19\textwidth]{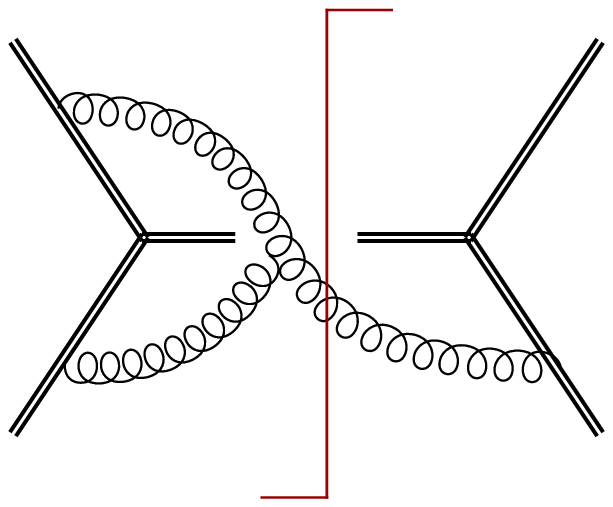} &&\includegraphics[height=0.19\textwidth]{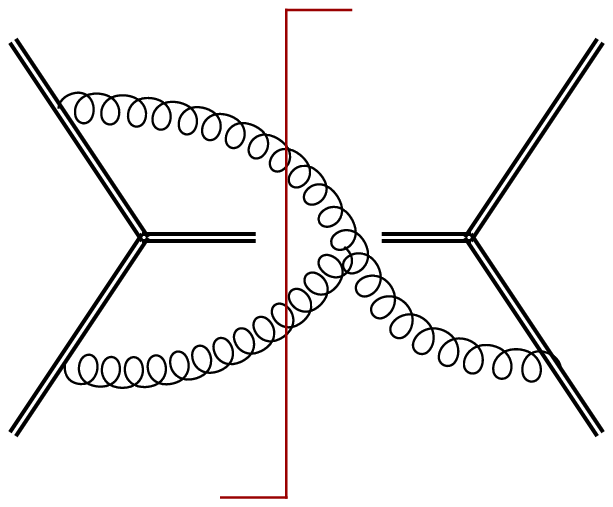}  && 
 \\[0.01cm]
  $D_{7,1}$&&$D_{7,2}$&&
  \end{tabular}
  \end{center}
   \vspace{-0.5cm}
\caption{Feynman diagrams that contribute to the soft function up to NNLO. In addition there are mirror symmetrical graphs, 
which we take into account by multiplying each diagram $D_i$ with a symmetry factor $f_i$, where $f_1=f_2=f_3=f_5=2$, $f_4=1$ and $f_6=f_7=4$. Additional diagrams, in which gluons attach to the jet Wilson line, vanish, see text. \label{fig:diagrams}}
\label{diag}
  \end{figure}

The diagrams contributing to the soft function to NNLO are shown in Figure \ref{fig:diagrams}. To understand why the color structure is diagonal, let us  work out the color factor of the diagram $D_1$ in the color-space formalism. The color structure associated with this gluon exchange is $-\bm{T}_1\cdot \bm{T}_2=-\sum_a  \bm{T}_1^a \bm{T}_2^a$. The minus sign arises, because one of the Wilson lines is conjugated. Its matrix element can be simplified as follows
\begin{align}
 \langle\, 1' \;2'\; J'\, | \,\left(- \bm{T}_1\cdot \bm{T}_2\right) \, |\, 1\; 2\; J\, \rangle_{\rm color} &=
  \langle\, 1' \;2'\; J'\, | \, \bm{T}_1\cdot \bm{T}_1 + \bm{T}_1\cdot \bm{T}_J \, | \,1\; 2\; J\, \rangle_{\rm color} \nonumber \\
 &= C_1\,   \langle\, 1'\; 2'\; J' | \,\bm{1}\, | 1\; 2\; J\, \rangle_{\rm color} + \langle\, 1' \;2'\; J'\, |  \bm{T}_1\cdot \bm{T}_J \, |\, 1\; 2\; J\, \rangle_{\rm color} \nonumber \\
    & = \frac{1}{2} \left( C_1 +C_2 -C_J \right) \langle\, 1'\; 2'\; J' \,| \,\bm{1}\, | 1\; 2\; J\,\, \rangle_{\rm color}\,,
\end{align}
where we have indicated with primes the different color state of the conjugate hard amplitude. 
In the first line, color conservation $\sum_i \bm{T}_i^a = 0$ was used. The Casimir operator in the second line is $ \bm{T}_1\cdot \bm{T}_1 = C_1 \bm{1}$. The third line follows after applying color conservation two more times. The color structure is thus trivial and it is convenient to define
\begin{equation}\label{casimir}
C_s = \frac{1}{2} \left( C_1 +C_2 -C_J \right) =
\begin{cases} \; C_F-C_A/2 & \text{ for } \; q\bar{q} \to g\,, \\
\; C_A/2 & \text{ for } \; q g \to q \;  \text{ and } \; gg\to g\,.
\end{cases}
\end{equation}
Using color conservation and the commutation relations $\left[\bm{T}_i^a,\bm{T}_i^b \right] = i f^{abc} \bm{T}_i^c$, the color factor of all diagrams can be expressed in terms of the Casimir invariants multiplying the unit matrix and the soft function takes the form
\begin{equation}
 \langle\, 1' \;2'\; J' | \, \bm{S}(\omega) \, | 1\; 2\; J\, \rangle_{\rm color} =   S(\omega) \langle\, 1'\; 2'\; J' |\,\bm{1}\, | 1\; 2\; J\, \rangle_{\rm color}\,.
\end{equation}
The color structure in the factorization theorem (\ref{factform}) is trivial and it can be written in terms of scalar hard, jet and soft functions.

\begin{figure}[t!]
 \begin{center}
 \begin{tabular}{ccccc}\
 \includegraphics[height=0.19\textwidth]{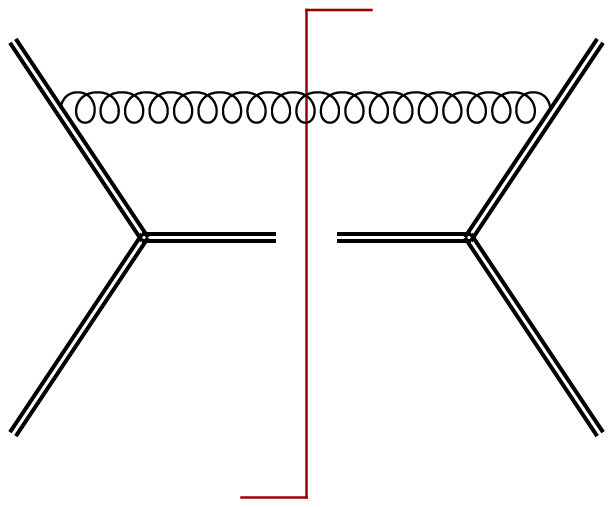} &&\includegraphics[height=0.19\textwidth]{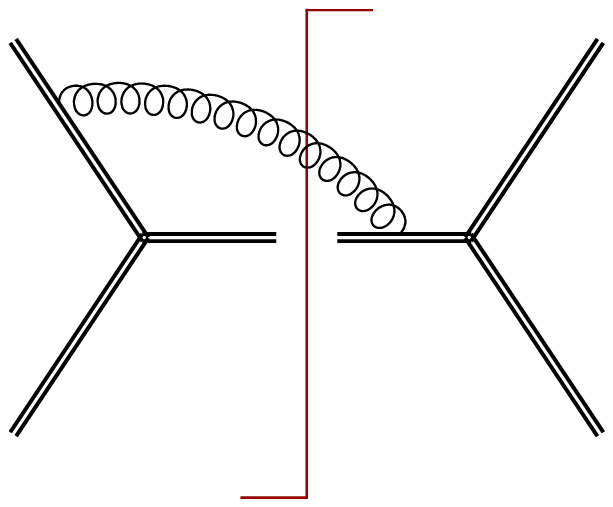}  &&\includegraphics[height=0.19\textwidth]{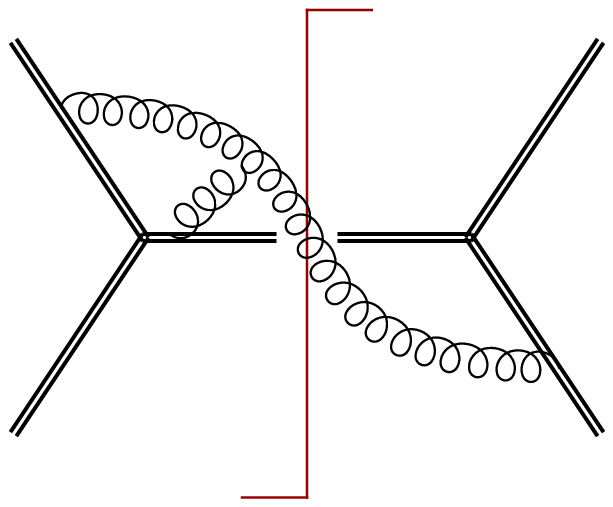}\\[0.1cm]
 $(a)$&&$(b)$&&$(c)$ 
  \end{tabular}
  \end{center}
  \vspace{-0.5cm}
  \caption{Additional diagrams that turn out to be zero. Graphs (a) and (b) vanish because the associated integrals are scaleless, (c) because of its color structure.}
  \label{fig:zero}
  \end{figure}

Not only the color structure, but also the dependence of the soft function on the light-cone vectors $n_i$ is very simple, because of the invariance of the Wilson lines under a rescaling $n_i \to \lambda \, n_i$ of the reference vectors. From the definition (\ref{softdef}) it follows that under a simultaneous rescaling
\begin{align}\label{rescale}
n_1 &\to \lambda_1\, n_1\,, & n_2 &\to \lambda_2\, n_2\,, & n_J &\to \lambda_J \, n_J \,, &  
\omega &\to \lambda_J \, \omega 
\end{align}
 the soft function transforms as
\begin{equation}
 S(\omega) \to \frac{1}{\lambda_J}  S(\omega)\,.
\end{equation}
The dependence of the soft function on the reference vectors must thus be of the form
\begin{equation}\label{omegahat}
S(\omega) = \frac{1}{\omega} \; f( \hat{\omega}) \;\;\; \text{ with }\;\; \hat{\omega} = \omega\, \sqrt{\frac{2\, n_1\cdot n_2}{n_1\cdot n_J \; n_2\cdot n_J}}\,,
\end{equation}
where the quantity $\hat{\omega}$ is invariant under the rescaling (\ref{rescale}). The factor of $\sqrt{2}$ in $\hat{\omega}$ has been inserted for convenience.

Having discussed its general structure, we now turn to the evaluation of the soft function to NNLO.  
Since the soft Lagrangian of SCET is identical to the ordinary QCD Lagrangian, the computation can be performed in QCD itself. At tree level only the vacuum state contributes, and the soft function is trivially given by a delta-function. At NLO both virtual corrections to the vacuum state and one-gluon emission diagrams arise. In dimensional regularization the virtual graphs are scaleless and vanish. As the kinematic dependence of the soft function involves all of the reference vectors $n_i$, see (\ref{omegahat}), the diagrams must have at least one gluon attachment to the Wilson line $\bm{S}_1$ and one to $\bm{S}_2$. It follows that only the diagram $D_1$ in Figure \ref{fig:diagrams} gives a non-vanishing contribution at NLO, while diagrams such as (a) and (b) in Figure \ref{fig:zero} are scaleless and vanish. The calculation of diagram $D_1$ reduces to the computation of the integral
\begin{equation}
I_1=\int d^dk \; \delta(k^2)\,\theta(k^0)\;
\frac{n_1\cdot n_2}{n_1\cdot k\; n_2\cdot k}\;
\delta(\omega-n_J\cdot k)\,.
\end{equation}
This integral has been evaluated in \cite{Becher:2009th} in an expansion in $\eps = (4-d)/2$ up to the finite term, but here we also need higher order terms. As detailed in Appendix \ref{oneloop}, we have computed this integral in closed form and obtain
\begin{equation}
I_1= - \frac{2\pi^{1-\eps}}{\omega} \;  \hat{\omega}^{-2\eps} \,\frac{\Gamma(\eps)\,\Gamma^2(1-\eps)}{\Gamma(1-2\eps)}\,,
\end{equation}
which is in agreement with the result from \cite{Becher:2009th}.

At NNLO the purely virtual corrections are again scaleless and vanish. Among the mixed virtual-real and the double real emissions, only the diagrams in Figure \ref{fig:diagrams} are non-zero. The respective color factors can be worked out in the color-space formalism, as discussed above. We find $C_s$ for diagram $D_2$, $C_s^2$ for diagram $D_3$, $C_s(C_s-C_A/2)$ for diagrams $D_4$ and $D_5$ and $C_s C_A/2$ for the non-abelian diagrams $D_6$ and $D_7$. Note that the diagrams in Figure \ref{fig:diagrams} only involve attachments to the Wilson lines $\bm{S}_1$ and $\bm{S}_2$. In addition there are diagrams with attachments to the Wilson line  $\bm{S}_J$. An example of such a diagram is the graph (c) in Figure \ref{fig:zero}. The color structure of this diagram is
 \begin{align}\label{antisymm}
 f_{abc}\; \bm{T}_1^a \,\bm{T}_2^b \,\bm{T}_J^c \,,
\end{align}
which, by color conservation, vanishes when acting on three-parton states $| 1\; 2\; J\, \rangle_{\rm color}$. In fact, the diagram vanishes in general: it could only be non-zero if it contained an anti-symmetric function of the three vectors $n_1$, $n_2$ and $n_J$. However, the only invariant $\hat{\omega}$ is symmetric under the interchange $n_1\leftrightarrow n_2$ and thus all diagrams with the color structure (\ref{antisymm}) vanish. 
The remaining abelian diagrams with attachments to all three Wilson lines also vanish, because at least one of the subdiagrams involves only two light-cone vectors and these integrals are thus scaleless.

The technical aspects of the NNLO calculation are given in Appendix \ref{twoloop}, where we discuss the evaluation and give results for all integrals associated with the diagrams in Figure \ref{fig:diagrams}. After combining all contributions, we obtain the following result for the bare soft function
\begin{multline}
   S^{\rm bare}(\omega) =  \delta(\omega) + \frac{Z_\alpha \alpha_s}{(4\pi)}\, \frac{1}{\omega} \left(\frac{\mu}{\hat{\omega}}\right)^{2\eps} C_s\, S_1(\eps)     \\ 
   +  \frac{Z_\alpha^2 \alpha_s^2}{(4\pi)^2} \,\frac{1}{\omega} \left(\frac{\mu}{\hat{\omega}}\right)^{4\eps}\, C_s\, \left[   C_A
 S_{A}(\eps) +  n_f T_F S_{f}(\eps) +  C_s S_{s}(\eps)  \right]\,,
   \end{multline}
where $\alpha_s$ refers to the $\overline{\rm MS}$ coupling constant, which is related to the bare coupling constant $\alpha_s^0$ via  $Z_\alpha\, \alpha_s \,\mu^{2\eps} = e^{-\eps \gamma_E}(4\pi)^\eps \alpha_s^0$ with $Z_\alpha  = 1-\beta_0 \alpha_s/(4\pi \eps)$ and $\beta_0 = 11/3 \,C_A - 4/3\, T_F n_f$. The variable $\hat\omega$ includes the dependence on the light-cone vectors and was defined in (\ref{omegahat}). The one-loop coefficient reads
   \begin{equation}
 S_1(\eps) = - 8 \,e^{\gamma_E \eps } \,\frac{\Gamma(\eps)\Gamma^2(1-\eps)}{\Gamma(1-2\eps)}\,.
   \end{equation}
The three leading terms in $\epsilon$ were given in \cite{Becher:2009th}. The two-loop coefficients are found to be
 \begin{align}
 S_{A}(\epsilon) =& -\frac{44}{3 \epsilon ^2}+\frac{1}{\epsilon }\left(\frac{4 \pi^2}{3}-\frac{268}{9}\right)+\frac{22 \pi
   ^2}{3}-\frac{1616}{27}+56 \zeta _3 \nonumber \\
   &+\left(-\frac{9712}{81} +\frac{134 \pi^2}{9}+ \frac{2728 \zeta _3}{9}+\frac{88 \pi ^4}{45} \right)\, \epsilon  
    + \mathcal{O}(\eps^2)\, ,\nonumber\\
    S_{f}(\epsilon) =& \frac{16}{3 \epsilon ^2}+\frac{80}{9 \epsilon }-\frac{8 \pi
   ^2}{3}+\frac{448}{27} +\left(\frac{2624}{81}-\frac{40 \pi
   ^2}{9}-\frac{992 \zeta _3}{9}\right)\,\epsilon + \mathcal{O}(\eps^2)\, , \nonumber\\
   S_{s}(\epsilon) =& -\frac{32}{\epsilon ^3}+\frac{80 \pi ^2}{3
   \epsilon }+\frac{1984 \zeta _3}{3}+\frac{100 \pi^4}{9} \,\epsilon  + \mathcal{O}(\eps^2)\, . 
   \end{align}
The function $S^{\rm bare}(\omega)$ is a distribution in $\omega$ whose explicit form is obtained after expanding
 \begin{equation} \label{eq:star}
\frac{1}{\omega} \left(\frac{\omega}{\mu}\right)^{-n\epsilon} = -\frac{1}{n\epsilon} \delta(\omega) + \sum_{m=0}^{\infty} \frac{(-n\epsilon)^m}{m!} \left[ \frac{\ln^m(\frac{\omega}{\mu})}{\omega} \right]^{[\mu]}_* \,.
 \end{equation}
The star-distributions are generalizations of plus-distributions to dimensionful variables, their definition can be found in \cite{Bosch:2004th}. Because the expansion (\ref{eq:star}) starts at $1/\epsilon$, we evaluated the two-loop coefficients to $\mathcal{O}(\epsilon)$.

\section{Renormalization\label{renorm}}

The renormalization of the soft function is conveniently discussed in Laplace space. We introduce the Laplace transformed soft function
\begin{equation}
\widetilde{s}(L,\mu) = \int_0^\infty d \omega\, e^{-\nu \omega}\, S(\omega,\mu)\;\;\text{ with }\;  \nu= \frac{1}{\kappa \,e^{\gamma_E}}\,,
\end{equation}
where, for later convenience, we have written $\widetilde{s}$ as a function of the logarithm of the Laplace-space variable
\begin{equation}
L= \ln\frac{\hat{\kappa}}{\mu} = 
\ln
\left(\frac{\kappa}{\mu}\sqrt{\frac{2n_1\cdot n_2}{n_1\cdot n_J\;n_2\cdot n_J}}\right)\,.
\end{equation}
The Laplace transform of the bare function is immediately obtained using the relation
\begin{equation}
\int_0^\infty d \omega\, e^{-\nu \omega}\, \omega^{-1-n\epsilon } = e^{-n \epsilon\, \gamma_E}\, \Gamma(-n\epsilon)\, \kappa^{-n\, \epsilon}\,.
\end{equation}
After expanding in $\epsilon$, the Laplace transformed function is a polynomial in $L$. The function $\widetilde{s}$ is also what is needed to perform soft-gluon resummation in the momentum space formalism of \cite{Becher:2006nr}. It fulfils the renormalization group (RG) equation 
\begin{align} \label{eq:RG}
\frac{\rd}{\rd \ln\mu}  \, \widetilde{s}\left(L,\mu\right)
&= \left[ - 4 \,\Gamma_{\mathrm{cusp}}\, \ln\frac{\hat{\kappa}}{\mu}    -2 \gamma^{S} \right] \widetilde{s} \left(L,\mu\right) \,,
\end{align}
where $\Gamma_{\mathrm{cusp}}$ denotes the cusp anomalous dimension, which is known at the three-loop level \cite{Moch:2004pa}. The anomalous dimension $\gamma^{S}$ was inferred to three loops in \cite{Becher:2009th} using RG invariance of the direct photon-production cross section and the three-loop results for the hard anomalous dimensions \cite{Becher:2009cu,Becher:2009qa}, the quark jet anomalous dimension \cite{Becher:2006mr} and the Casimir scaling property of the soft anomalous dimension. Expanding  the anomalous dimensions as 
$\Gamma_{\mathrm{cusp}}= \sum_{n=0}^\infty \Gamma_n (\frac{\alpha_s}{4\pi})^{n+1}$ and 
$\gamma^{S} = \sum_{n=0}^\infty \gamma^{S}_n \,(\frac{\alpha_s}{4\pi})^{n+1}$, one can easily solve the RG equation (\ref{eq:RG}). To two-loop order, the solution takes the form
\begin{multline}
\widetilde{s} \left( L ,\mu \right) = 
1 + \left( \frac{\alpha_s}{4 \pi} \right)
\left[ 2\Gamma_0 L^2 + 2\gamma_0^S L + c_1^S \right] 
+\left( \frac{\alpha_s}{4 \pi} \right)^2 
\bigg[ 2 \Gamma_0^2 L^4 
- \frac{4\Gamma_0}{3} \left( \beta_0 - 3 \gamma_0^S \right) 
L^3 \\  
 + 2 \left( \Gamma_1 + (\gamma_0^S)^2 - \beta_0 \gamma_0^S + \Gamma_0 c_1^S \right) L^2 + 2 \left (\gamma_1^S + \gamma_0^S c_1^S -\beta_0 c_1^S\right) L + c_2^S \bigg] \,.
\end{multline}
The expansion coefficients of the anomalous dimensions are
\begin{align}
\Gamma_0 &= 4C_s \,, &
\Gamma_1 &= 4C_s \left[  C_A \left( \frac{67}{9} - \frac{\pi^2}{3} \right) 
- \frac{20}{9}T_F n_f   \right]\,, \\
\gamma_0^{S} &= 0 \,, & 
\gamma_1^{S} &= C_s \,C_A\left( 28 \zeta_3 - \frac{808}{27} + \frac{11 \pi^2}{9} \right)  +C_s\, n_f T_F \left(
\frac{224}{27} - \frac{4 \pi^2}{9} \right)  \nonumber \, ,
\end{align}
and the one-loop constant is $c^{S}_1=C_s\, \pi^2$. The color factor $C_s$ depends on the partonic channel and was defined in (\ref{casimir}).

The function $\widetilde{s}$ renormalizes multiplicatively, $\widetilde{s} = Z_{s} \,\widetilde{s}^{\rm bare}$, and $Z_{s}$ fulfils the same RG equation (\ref{eq:RG}) as the renormalized soft function. Solving this equation \cite{Becher:2009cu,Becher:2009qa}, one derives the following expression for the logarithm of the $Z$-factor,
\begin{align} 
\ln{ Z}_{ s} &= \frac{\alpha _s}{4 \pi } \left[-\frac{\Gamma_0 }{\epsilon ^2}
+\frac{1}{
   \epsilon }\left(2 \Gamma_0\, L+\gamma^{S}_0\right)\right]
 \nonumber\\
&\quad
+ \left(\frac{\alpha _s}{4 \pi }\right)^2
  \Bigg[  \frac{3 \beta _0 \Gamma_0 }{4 \epsilon ^3} 
 -\frac{\beta _0}{2 \epsilon ^2}\left(2  \Gamma_0  L + \gamma^{S}_0 \right) 
  -\frac{\Gamma_1 }{4\epsilon ^2} 
+\frac{1}{2 \epsilon}\left(2 \Gamma_1  L + \gamma^{S}_1 \right)
 \Bigg] \,. 
   \end{align}
Since all the necessary anomalous dimensions are known, the $Z$-factor is completely determined at the two-loop level. The cancellation of all divergences $1/\epsilon^n$, for $n=1\dots 4$, in the renormalized result provides a strong check of our calculation. We finally obtain for the non-logarithmic two-loop coefficient 
 \begin{equation}  
 c^{S}_2 =  C_s^2 \frac{\pi ^4 }{2} +
 C_s C_A  \left(\frac{2428}{81}+\frac{335 \pi^2}{54}-\frac{22 \zeta_3}{9}-\frac{14 \pi ^4}{15}\right)
 +C_s n_f T_F \left(-\frac{656}{81}-\frac{50 \pi ^2}{27}+\frac{8 \zeta_3}{9}\right)  \,,
\end{equation}
which is the main result of our paper. Note that the coefficient of the color structure $C_s^2$ is just one half of the one-loop coefficient squared. This is a consequence of the non-abelian exponentiation theorem for Wilson lines. Note that the relation would be more complicated for the momentum space function. Since the position-space and Laplace-space functions are related by Wick rotation, the non-abelian exponentiation takes the same form as in position space. 

To illustrate the size of the corrections, we now evaluate the soft function numerically for $n_f=5$. For $q\bar{q}\to g V$, where $C_s=C_F-C_A/2$, we find
\begin{multline}
\widetilde{s} \left( L ,\mu \right) = 1 + \left(\frac{\alpha_s}{4\pi}\right) \left(-1.333 L^2-1.645\right)  \\
+ \left(\frac{\alpha_s}{4\pi}\right)^2 \left(0.889 L^4+6.815 L^3-7.018 L^2+6.17 L+13.22\right)\,,
\end{multline}
while the case $C_s=C_A/2$, which is relevant for  $qg\to q V$ and for Higgs production, yields
\begin{multline}
\widetilde{s} \left( L,\mu \right) = 1 +\left(\frac{\alpha_s}{4\pi}\right) \left(12 L^2+14.80\right) \\
+ \left(\frac{\alpha_s}{4\pi}\right)^2 \left(72 L^4-61.33 L^3+260.6 L^2-55.53
   L+2.757\right)\,.
\end{multline}
In the first case, the corrections are quite small: setting $\alpha_s=0.1$ and varying $-2<L<2$ the two-loop contribution is below half a per-cent. The situation is somewhat different for $C_s=C_A/2$. While the constant piece is small, the corrections can be sizable, if the soft function is not evaluated at its natural scale: for $\alpha_s=0.1$ and $L=-2$, for example, the perturbative expansion reads $\widetilde{s} \left( L,\mu \right) = 1 + 0.50 +0.18$. In a resummed computation, the hard, jet and soft functions are evaluated near their natural scale and so we do not expect large corrections to the results obtained in \cite{Becher:2011fc}. However, to make definite statements, the different two-loop ingredients should be combined, which will be done elsewhere.

\section{Conclusions\label{concl}}

We have computed the soft emissions for electroweak boson production at large transverse momentum near the partonic threshold to two-loop order. The relevant soft function involves light-like Wilson lines in three directions: along the beam directions and along the direction of the final-state jet recoiling against the electroweak boson. The invariance of the Wilson lines under a rescaling of the reference vectors puts strong constraints on the form of the result and greatly simplifies the computation. In particular, the diagrams which involve emissions or absorptions from the jet Wilson line vanish in dimensional regularization, such that the computation ends up being similar to earlier computations of dijet soft functions. 

With our result, the last remaining ingredient to extend the threshold resummation for electroweak boson production processes to N$^3$LL is now available. In the future, we will combine our result with the two-loop jet and hard functions to obtain improved predictions for transverse momentum spectra. 

\vspace{0.6cm}
{\em Acknowledgments:\/}
This work is supported in part by funds provided by the Schweizerischer Nationalfonds (SNF). The Albert Einstein Center for Fundamental Physics at the University of Bern is supported by the Innovations- und Kooperationsprojekt C-13 of the Schweizerische Universit\"atskonferenz (SUK/CRUS).


\begin{appendix}

\section{Evaluation of the NLO integral\label{oneloop}}

In this appendix we outline the calculation of the NLO integral
\begin{equation}
I_1=\int d^dk \; \delta(k^2)\,\theta(k^0)\;
\frac{n_1\cdot n_2}{n_1\cdot k\; n_2\cdot k}\;
\delta(\omega-n_J\cdot k)\,.
\end{equation}
We closely follow the strategy adopted in \cite{Becher:2009th}. First we introduce light-cone coordinates and decompose the four-vectors as
\begin{equation}
q^\mu=q_+ \frac{n_1^\mu}{\sqrt{2n_1\cdot n_2}}+q_-\frac{n_2^\mu}{\sqrt{2n_1\cdot n_2}}+q_\perp^\mu
\end{equation}
with $n_i^2=0$ and $n_i\cdot q_\perp=0$. The result for the integral is frame independent, but it is convenient to work in a reference frame where $\vec{n}_1$ and $\vec{n}_2$ are back-to-back. Notice that this does not imply $n_1 \cdot n_2 = 2$. We may, for instance, choose a frame where $n_1 = (1,0,0,1)$ and $n_2 = (\lambda,0,0,-\lambda)$ with $\lambda = n_1 \cdot n_2/2$. In light-cone coordinates the integral takes the form
\begin{align}
I_1& = \Omega_{d-3} \int dk_+ dk_-
\int_{0}^{\infty} d|\vec{k}_\perp| \, |\vec{k}_\perp|^{d-3}
\int_{-1}^1 d\cos\theta \, \sin^{d-5}\theta 
\nonumber\\&\qquad\times
\delta(k_+ k_- -\vec{k}_\perp^2)\,\theta(k^0)\;\frac{1}{k_+ k_-}\;
\delta\Big(\omega-\frac{k_+ n_{J-}}{2}-\frac{k_ - n_{J+}}{2}+
\sqrt{n_{J+}n_{J-}} \,|\vec{k}_\perp| \cos\theta\Big)\,,
\end{align}
where $\theta$ is the angle between $\vec{k}_\perp$ and $\vec{n}_{J\perp}$ and $\Omega_n$ is the $n$-dimensional solid angle. We next use the on-shell condition to perform the $|\vec{k}_\perp|$-integration. Parameterizing 
\begin{equation}
k_+ = \frac{2\omega}{n_{J-}}\,x y, 
\qquad\qquad 
k_- = \frac{2\omega}{n_{J+}}\,(1-x)y\,,
\end{equation}
we arrive at
\begin{align}
I_1& = \frac{\Omega_{d-3}}{2\omega}  \; \hat{\omega}^{-2\eps}
\int_0^1 dx \int_0^\infty dy\;
(x \bar x)^{-1-\eps} \,y^{-1-2\eps}
\int_{-1}^1 d\cos\theta \, \sin^{d-5}\theta \;\;
\delta(1-y + 2 y \sqrt{x \bar x}  \cos\theta)
\end{align}
with $\bar x=1-x$ and $\hat{\omega}$ as defined in (\ref{omegahat}). We then use the delta-constraint to perform the $y$-integration. The subsequent integration over $\cos\theta$ gives rise to an hypergeometric function and one is left with
\begin{align}
I_1& = \frac{\pi^{1-\eps}}{\Gamma(1-\eps)} \,\frac{1}{\omega}  \; \hat{\omega}^{-2\eps}
\int_0^1 dx \; (x \bar x)^{-1-\eps} \;
{}_2F_{1}\left(\frac12-\epsilon,-\epsilon;1-\epsilon;4x \bar x\right)\,.
\end{align}
After performing a variable transformation to the argument of the hypergeometric function, the integral can be solved in closed form and we obtain
\begin{equation}
I_1= - \frac{2\pi^{1-\eps}}{\omega} \;  \hat{\omega}^{-2\eps} \,\frac{\Gamma(\eps)\,\Gamma^2(1-\eps)}{\Gamma(1-2\eps)}\,.
\end{equation}

\section{Details of the NNLO calculation\label{twoloop}}

The calculation of the NNLO diagrams proceeds along the same lines as the NLO calculation that we outlined in Appendix \ref{oneloop}. Here we briefly describe the new elements that arise at NNLO, and give results for the individual integrals associated with the diagrams in Figure \ref{fig:diagrams}.

We start with diagram $D_2$. As the soft function only depends on the total momentum of the radiation, it is convenient to express this diagram through the cut of the vacuum polarization
\begin{equation}
\Pi_{\mu\nu}(k) = (k^2\,g_{\mu\nu}-k_\mu k_\nu)\; \Pi(k^2)\,.
\end{equation}
Diagram $D_2$ then involves an integral of the form
\begin{align} 
\int d^dk \; 
\frac{n_1^\mu \,n_2^\nu}{n_1\cdot k\; n_2\cdot k\; (k^2)^2}\;
\delta(\omega-n_J\cdot k)\;\,\text{Im}\,\Pi_{\mu\nu}(k)\,.
\end{align}
The contribution from $k_\mu k_\nu$ vanishes, since the light-cone propagators cancel out in this term and one is left with a scaleless integral. Inserting the one-loop expression for the vacuum polarization, we obtain the integral
\begin{align} 
I_2=\int d^dk \; \theta(k^2)\,\theta(k^0)\;
\frac{n_1\cdot n_2}{n_1\cdot k\; n_2\cdot k\; (k^2)^{1+\eps}}\;
\delta(\omega-n_J\cdot k)\,.
\end{align}
The calculation of this integral is similar to the one sketched in Appendix \ref{oneloop}, except that we now use a slightly different parameterization given that $k^2\neq 0$, 
\begin{equation} \label{substitute}
k_+ = \frac{\omega}{n_{J-}}\,x y, 
\qquad\quad 
k_- = \frac{\omega}{n_{J+}}\,(1-x)y\,,
\qquad\quad 
|\vec{k}_\perp| = \frac{\omega}{\sqrt{n_{J+}n_{J-}}}\sqrt{(1-x)x}\;y\, u\, .
\end{equation}
We then use the delta-constraint to perform the $y$-integration. The remaining integrals are standard and can be performed along the same lines as in Appendix \ref{oneloop}, yielding
\begin{align} 
I_2= \frac{4\pi^{1-\eps}}{\omega} \;  \hat{\omega}^{-4\eps} \,
\frac{\Gamma(1-\eps)\,\Gamma(1+2\eps)\,\Gamma^2(-2\eps)}{\Gamma(1-4\eps)}\,.
\end{align}

Next, we turn to the evaluation of the abelian diagrams $D_3$, $D_4$ and $D_5$. The planar diagram $D_3$ involves the integral
\begin{equation}
I_3=\int [dk]\, \int [dl]\;\,
\frac{(n_1\cdot n_2)^2\;\delta(\omega-n_J\cdot k-n_J\cdot l)}
{n_1\cdot k\; (n_1\cdot k + n_1\cdot l)\;
n_2\cdot l\;(n_2\cdot k+n_2\cdot l)},
\end{equation}
where we introduced the short-hand notation $[dk] = d^dk \; \delta(k^2)\,\theta(k^0)$.
Using partial fractioning identities this integral can be expressed as $I_3 = I_4/2 - I_5$, where $I_4$ and $I_5$ are integrals associated with diagrams $D_4$ and $D_5$. Their precise definition will be given below. The integral in diagram $D_4$,
\begin{equation}
I_4=\int [dk]\, \int [dl]\;\,
\frac{(n_1\cdot n_2)^2\;\delta(\omega-n_J\cdot k-n_J\cdot l)}
{n_1\cdot k\; n_1\cdot l\;
n_2\cdot k\;n_2\cdot l}\,,
\end{equation}
can easily be calculated since it is just the convolution of two NLO integrals,
\begin{align}
I_4&=\int_0^\omega d\omega'\;
I_1(\omega-\omega')\,I_1(\omega')
= \frac{\pi^{2-2\eps}}{\omega} \;  \hat{\omega}^{-4\eps} \,\frac{\Gamma^2(1+\eps)\,\Gamma^4(-\eps)}{\Gamma(-4\eps)}\,.
\end{align}
The non-planar diagram $D_5$ is more complicated. It gives rise to the integral
\begin{equation}
I_5=\int [dk]\, \int [dl]\;\,
\frac{(n_1\cdot n_2)^2\;\delta(\omega-n_J\cdot k-n_J\cdot l)}
{n_1\cdot k\; (n_1\cdot k + n_1\cdot l)\;
n_2\cdot k\;(n_2\cdot k+n_2\cdot l)}\,.
\end{equation}
As the delta-function only constraints the sum of the cut momenta, we first combine them to $q=k+l$ and rewrite the integral as
\begin{equation}
I_5=\int d^d q\;
\frac{(n_1\cdot n_2)\;\delta(\omega-n_J\cdot q)}
{n_1\cdot q\;n_2\cdot q}\;
\int [dk]\, \int [dl]\;\,
\frac{(n_1\cdot n_2)\;\delta^d(q-k-l)}
{n_1\cdot k\; n_2\cdot k}\,.
\end{equation}
The integrations over $k$ and $l$ can be performed using the auxiliary integral (\ref{Iaux3}) from Appendix \ref{aux}. We further adopt the parameterization (\ref{substitute}) together with an integral representation of the hypergeometric function and arrive at
\begin{align}
I_5& = 2^{1+4\eps} \,\pi^{1-\eps}\,\frac{\Omega_{d-3}}{\omega}  \; \hat{\omega}^{-4\eps}\,
\frac{\Gamma(1-\eps)}{\Gamma(1-2\eps)}\,
\int_0^1 dx \int_0^\infty dy\;
(x \bar x)^{-1-2\eps} \,y^{-1-4\eps}
\int_{-1}^1 d\cos\theta \, \sin^{d-5}\theta \;
\nonumber\\&\quad\times
\int_0^1 du\;\int_0^1 dv\;\;
u^{1-2\eps} \,(1-u^2)^{-1-2\eps}\;
v^{-1-\eps} \,(1-v u^2)^{\eps}\;\;
\delta\Big(1-\frac{y}{2} + y \sqrt{x \bar x} \,u \cos\theta\Big)\,.
\end{align}
We now use the delta-constraint to perform the $y$-integration. The subsequent integrations over $\cos\theta$, $x$ and $v$ are standard and can be performed along the same lines as in Appendix \ref{oneloop}. This yields an integral over a product of two hypergeometric functions,
which we solve order by order in the $\eps$-expansion. The result reads
\begin{align}
I_5
= -\frac{\pi^{2-2\eps}}{\omega} \;  \hat{\omega}^{-4\eps} \;
\frac{\Gamma^2(1-\eps)}{\Gamma^2(1-2\eps)}\,
\left( \frac{1}{\eps^3}-8\zeta_3-\frac{2\pi^4}{5} \eps + \mathcal{O}(\eps^2)
\right).
\end{align}

We finally turn to the non-abelian diagrams $D_6$ and $D_7$. For the integral associated with diagram $D_6$,
\begin{align}
I_6&=\int [dk]\, \int [dl]\;\,
\frac{(n_1\cdot n_2)\;(n_1\cdot l - n_1\cdot k)\;\delta(\omega-n_J\cdot k-n_J\cdot l)}
{n_1\cdot k\; (n_1\cdot k + n_1\cdot l)\;
\;(n_2\cdot k+n_2\cdot l)\;(k+l)^2}\,,
\end{align}
we again combine the cut momenta to $q=k+l$ and make use of the auxiliary integrals from Appendix \ref{aux}. The subsequent steps are by now straightforward and we obtain a closed expression
\begin{align}
I_6&=
\frac{\pi^{2-2\eps}}{\omega} \;  \hat{\omega}^{-4\eps} \,
\frac{\Gamma(1+2\eps)\,\Gamma(-2\eps)\,\Gamma^2(-\eps)}{(1-2\eps)\Gamma(1-4\eps)}\,.
\end{align}
The last diagram $D_7$ contains both a one-particle and a two-particle cut. The one-particle cut involves the real part of the loop integral
\begin{align}
& i \int d^d k\;\,
\frac{n_1\cdot n_2}
{(n_1\cdot q + n_1\cdot k)\;n_2\cdot k\;
(q+k)^2\;k^2} 
\nonumber\\
&\qquad \qquad
=\;
2 \pi^{2-\eps} \;e^{-i\pi\eps} \;\frac{\Gamma^2(1+\eps)\Gamma^3(-\eps)}{\Gamma(-2\eps)} \left( \frac{n_1\cdot n_2}{2\,n_1\cdot q \;n_2\cdot q} \right)^{1+\eps},
\end{align}
which gives rise to
\begin{equation}
I_{7,1}=\int d^dq \; \delta(q^2)\,\theta(q^0)\;
\left( \frac{n_1\cdot n_2}{2\,n_1\cdot q\; n_2\cdot q}\right)^{1+\eps}
\delta(\omega-n_J\cdot q)\,.
\end{equation}
This integral is similar to the NLO integral that we discussed in detail in Appendix \ref{oneloop}. It can be computed along the same lines and gives
\begin{equation}
I_{7,1}=
\frac{\pi^{1-\eps}}{2\omega} \;  \hat{\omega}^{-4\eps} \,
\frac{\Gamma(1+3\eps)\,\Gamma^2(-2\eps)}{\Gamma(-4\eps)\,\Gamma^2(1+\eps)}\,.
\end{equation}
Finally, the two-particle cut of diagram $D_7$ leads to the integral
\begin{align}
I_{7,2}&=\int [dk]\, \int [dl]\;\,
\frac{(n_1\cdot n_2)\;(n_2\cdot k + 2 \,n_2\cdot l)\;\delta(\omega-n_J\cdot k-n_J\cdot l)}
{n_1\cdot k\; (n_2\cdot k + n_2\cdot l)\;
\;n_2\cdot l\;(k+l)^2}\,.
\end{align}
For this integral we follow the same strategy that we adopted for the integral $I_5$. We again encounter a product of two hypergeometric functions, which we solve order by order in the $\eps$-expansion. The final result reads
\begin{align}
I_{7,2}
= -\frac{2\pi^{2-2\eps}}{\omega} \;  \hat{\omega}^{-4\eps} \;
\frac{\Gamma^2(1-\eps)}{\Gamma^2(1-2\eps)}\,
\left( \frac{1}{\eps^3}-\frac{\pi^2}{4\eps}-\frac{41}{2}\zeta_3-\frac{5\pi^4}{8} \eps + \mathcal{O}(\eps^2)
\right).
\end{align}

\section{Auxiliary integrals}\label{aux}

For the evaluation of the diagrams with two-particle cuts, the following integrals are useful:
\begin{align}
&\int [dk]\, \int [dl]\;\, \delta^d(q-k-l) =
\pi^{1-\eps} \,\frac{\Gamma(1-\eps)}{2\Gamma(2-2\eps)}\;(q^2)^{-\epsilon},\\[0.2em]
&\int [dk]\, \int [dl]\;\, \frac{\delta^d(q-k-l)}{n_1\cdot k} =
\pi^{1-\eps} \,\frac{\Gamma(-\eps)}{2\Gamma(1-2\eps)}\;\frac{(q^2)^{-\epsilon}}{n_1\cdot q},\\[0.2em]
&\int [dk]\, \int [dl]\;\, \frac{(n_1\cdot n_2)\;\delta^d(q-k-l)}{2\,n_1\cdot k\;n_2\cdot k} 
\nonumber\\
&\qquad=
\pi^{1-\eps} \,\frac{\Gamma(-\eps)}{\Gamma(1-2\eps)}
\left(\frac{2 \,n_1\cdot q\; n_2 \cdot q}{n_1\cdot n_2}\right)^{\epsilon}\;(q^2)^{-1-2\epsilon}\;
{}_2F_1\left(-\epsilon,\,-\epsilon;\,1-\epsilon;\,\frac{(n_1\cdot n_2) \;|\vec{q}_\perp|^2}
{2\,n_1\cdot q\;n_2\cdot q}\right),\label{Iaux3}\\[0.2em]
&\int [dk]\, \int [dl]\;\, \frac{(n_1\cdot n_2)\;\delta^d(q-k-l)}{2\,n_1\cdot k\;n_2\cdot l} 
\nonumber\\
&\qquad=
\pi^{1-\eps} \,\frac{\Gamma(-\eps)}{\Gamma(1-2\eps)}
\left(\frac{2 \,n_1\cdot q\; n_2 \cdot q}{n_1\cdot n_2}\right)^{\epsilon}\;|\vec{q}_\perp|^{-2-2\epsilon}\;(q^2)^{-\epsilon}\;
{}_2F_1\left(-\epsilon,\,-\epsilon;\,1-\epsilon;\,\frac{(n_1\cdot n_2) \;q^2}
{2\,n_1\cdot q\;n_2\cdot q}\right),
\end{align}
where $[dk] = d^dk \; \delta(k^2)\,\theta(k^0)$. Note that the above relations imply $q^2\geq0$ and $q^0\geq0$, since the momentum $q$ is the sum of two physical momenta. Our results are in agreement with \cite{Li:2011zp}.

\end{appendix}


\begin{thebibliography}{99}  
 
 \bibitem{Melnikov:2006di}
 K.~Melnikov and F.~Petriello,
 Phys.\ Rev.\ Lett.\  {\bf 96}, 231803 (2006)
 [arXiv:hep-ph/0603182].

\bibitem{Melnikov:2006kv}
 K.~Melnikov and F.~Petriello,
 Phys.\ Rev.\  D {\bf 74}, 114017 (2006)
 [arXiv:hep-ph/0609070].

\bibitem{Catani:2009sm}
 S.~Catani, L.~Cieri, G.~Ferrera, D.~de Florian and M.~Grazzini,
 Phys.\ Rev.\ Lett.\  {\bf 103}, 082001 (2009)
 [arXiv:0903.2120 [hep-ph]].
 
\bibitem{Gavin:2010az} 
  R.~Gavin, Y.~Li, F.~Petriello and S.~Quackenbush,
  Comput.\ Phys.\ Commun.\  {\bf 182}, 2388 (2011)
  [arXiv:1011.3540 [hep-ph]].

\bibitem{Garland:2001tf}
 L.~W.~Garland, T.~Gehrmann, E.~W.~N.~Glover, A.~Koukoutsakis and E.~Remiddi,
 Nucl.\ Phys.\  B {\bf 627}, 107 (2002)
 [arXiv:hep-ph/0112081].

\bibitem{Garland:2002ak}
 L.~W.~Garland, T.~Gehrmann, E.~W.~N.~Glover, A.~Koukoutsakis and E.~Remiddi,
 Nucl.\ Phys.\  B {\bf 642}, 227 (2002)
 [arXiv:hep-ph/0206067].

\bibitem{Gehrmann:2011ab} 
  T.~Gehrmann and L.~Tancredi,
  arXiv:1112.1531 [hep-ph].

\bibitem{Gehrmann:2011aa}
 T.~Gehrmann, M.~Jaquier, E.~W.~N.~Glover and A.~Koukoutsakis,
 arXiv:1112.3554 [hep-ph].
 
 \bibitem{Bauer:2000yr}
C.~W.~Bauer, S.~Fleming, D.~Pirjol and I.~W.~Stewart,
Phys.\ Rev.\  D {\bf 63}, 114020 (2001)
[arXiv:hep-ph/0011336].

\bibitem{Bauer:2001yt}
C.~W.~Bauer, D.~Pirjol and I.~W.~Stewart,
Phys.\ Rev.\  D {\bf 65}, 054022 (2002)
[arXiv:hep-ph/0109045].

\bibitem{Beneke:2002ph}
M.~Beneke, A.~P.~Chapovsky, M.~Diehl and T.~Feldmann,
Nucl.\ Phys.\  B {\bf 643}, 431 (2002)
[arXiv:hep-ph/0206152].


\bibitem{Becher:2009th}
 T.~Becher and M.~D.~Schwartz,
 JHEP {\bf 1002}, 040 (2010)
 [arXiv:0911.0681 [hep-ph]].

\bibitem{Becher:2006qw}
 T.~Becher and M.~Neubert,
 Phys.\ Lett.\  B {\bf 637}, 251 (2006)
 [arXiv:hep-ph/0603140].

\bibitem{Becher:2010pd}
 T.~Becher and G.~Bell,
 Phys.\ Lett.\  B {\bf 695}, 252 (2011)
 [arXiv:1008.1936 [hep-ph]].

\bibitem{Kidonakis:1999ur}
 N.~Kidonakis and V.~Del Duca,
 Phys.\ Lett.\  B {\bf 480}, 87 (2000)
 [arXiv:hep-ph/9911460].

\bibitem{Kidonakis:2003xm}
 N.~Kidonakis and A.~Sabio Vera,
 JHEP {\bf 0402}, 027 (2004)
 [arXiv:hep-ph/0311266].

\bibitem{Gonsalves:2005ng}
 R.~J.~Gonsalves, N.~Kidonakis and A.~Sabio Vera,
 Phys.\ Rev.\ Lett.\  {\bf 95}, 222001 (2005)
 [arXiv:hep-ph/0507317].

\bibitem{Becher:2011fc}
 T.~Becher, C.~Lorentzen and M.~D.~Schwartz,
 arXiv:1106.4310 [hep-ph].

\bibitem{KidonakisNew}
 N.~Kidonakis and R.~J.~Gonsalves,
 arXiv:1201.5265 [hep-ph].
 
\bibitem{Becher:2009cu}
 T.~Becher and M.~Neubert,
 Phys.\ Rev.\ Lett.\  {\bf 102}, 162001 (2009)
 [arXiv:0901.0722 [hep-ph]].

\bibitem{Gardi:2009qi}
 E.~Gardi and L.~Magnea,
 JHEP {\bf 0903}, 079 (2009)
 [arXiv:0901.1091 [hep-ph]].

\bibitem{Becher:2009qa}
 T.~Becher and M.~Neubert,
 JHEP {\bf 0906}, 081 (2009)
 [arXiv:0903.1126 [hep-ph]].

\bibitem{Catani:1996jh}
 S.~Catani and M.~H.~Seymour,
 Phys.\ Lett.\  B {\bf 378}, 287 (1996)
 [arXiv:hep-ph/9602277].

\bibitem{Catani:1996vz}
 S.~Catani and M.~H.~Seymour,
 Nucl.\ Phys.\  B {\bf 485}, 291 (1997)
 [Erratum-ibid.\  B {\bf 510}, 503 (1998)]
 [arXiv:hep-ph/9605323].

\bibitem{Belitsky:1998tc}
 A.~V.~Belitsky,
 Phys.\ Lett.\  B {\bf 442}, 307 (1998)
 [arXiv:hep-ph/9808389].

\bibitem{Becher:2005pd}
 T.~Becher and M.~Neubert,
 Phys.\ Lett.\  B {\bf 633}, 739 (2006)
 [arXiv:hep-ph/0512208].

\bibitem{Kelley:2011ng}
 R.~Kelley, M.~D.~Schwartz, R.~M.~Schabinger and H.~X.~Zhu,
 Phys.\ Rev.\  D {\bf 84}, 045022 (2011)
 [arXiv:1105.3676 [hep-ph]].

\bibitem{Monni:2011gb}
 P.~F.~Monni, T.~Gehrmann and G.~Luisoni,
 JHEP {\bf 1108}, 010 (2011)
 [arXiv:1105.4560 [hep-ph]].
 
\bibitem{Hornig:2011iu} 
  A.~Hornig, C.~Lee, I.~W.~Stewart, J.~R.~Walsh and S.~Zuberi,
  JHEP {\bf 1108}, 054 (2011)
  [arXiv:1105.4628 [hep-ph]].

\bibitem{Li:2011zp}
 Y.~Li, S.~Mantry and F.~Petriello,
 Phys.\ Rev.\  D {\bf 84}, 094014 (2011)
 [arXiv:1105.5171 [hep-ph]].

\bibitem{Kelley:2011aa}
 R.~Kelley, M.~D.~Schwartz, R.~M.~Schabinger and H.~X.~Zhu,
 arXiv:1112.3343 [hep-ph].

\bibitem{Aybat:2006wq}
 S.~M.~Aybat, L.~J.~Dixon and G.~F.~Sterman,
 Phys.\ Rev.\ Lett.\  {\bf 97}, 072001 (2006)
 [arXiv:hep-ph/0606254].

\bibitem{Aybat:2006mz}
 S.~M.~Aybat, L.~J.~Dixon and G.~F.~Sterman,
 Phys.\ Rev.\  D {\bf 74}, 074004 (2006)
 [arXiv:hep-ph/0607309].

\bibitem{Mitov:2009sv}
 A.~Mitov, G.~F.~Sterman and I.~Sung,
 Phys.\ Rev.\  D {\bf 79}, 094015 (2009)
 [arXiv:0903.3241 [hep-ph]].

\bibitem{Ferroglia:2009ep}
 A.~Ferroglia, M.~Neubert, B.~D.~Pecjak and L.~L.~Yang,
 Phys.\ Rev.\ Lett.\  {\bf 103}, 201601 (2009)
 [arXiv:0907.4791 [hep-ph]].

\bibitem{Ferroglia:2009ii}
 A.~Ferroglia, M.~Neubert, B.~D.~Pecjak and L.~L.~Yang,
 JHEP {\bf 0911}, 062 (2009)
 [arXiv:0908.3676 [hep-ph]].

\bibitem{Dixon:2009ur}
 L.~J.~Dixon, E.~Gardi and L.~Magnea,
 JHEP {\bf 1002}, 081 (2010)
 [arXiv:0910.3653 [hep-ph]].

\bibitem{DelDuca:2011xm}
 V.~Del Duca, C.~Duhr, E.~Gardi, L.~Magnea and C.~D.~White,
 arXiv:1108.5947 [hep-ph].

\bibitem{DelDuca:2011ae}
 V.~Del Duca, C.~Duhr, E.~Gardi, L.~Magnea and C.~D.~White,
 arXiv:1109.3581 [hep-ph].

\bibitem{Bosch:2004th}
 S.~W.~Bosch, B.~O.~Lange, M.~Neubert and G.~Paz,
 Nucl.\ Phys.\  B {\bf 699}, 335 (2004)
 [arXiv:hep-ph/0402094].

\bibitem{Becher:2006nr}
 T.~Becher and M.~Neubert,
 Phys.\ Rev.\ Lett.\  {\bf 97}, 082001 (2006)
 [arXiv:hep-ph/0605050].

\bibitem{Moch:2004pa}
 S.~Moch, J.~A.~M.~Vermaseren and A.~Vogt,
 Nucl.\ Phys.\  B {\bf 688}, 101 (2004)
 [arXiv:hep-ph/0403192].

\bibitem{Becher:2006mr}
 T.~Becher, M.~Neubert and B.~D.~Pecjak,
 JHEP {\bf 0701}, 076 (2007)
 [arXiv:hep-ph/0607228].
 
 
  \end{thebibliography}
\end{document}